\documentclass[12pt]{article}
\usepackage{amsmath}
\usepackage{amssymb}
\usepackage{amsthm}
\usepackage{graphicx}

\usepackage{multicol}
\usepackage{xcolor}
\usepackage{cite}
\usepackage{subcaption}
\usepackage{enumerate}
\usepackage{hyperref}

\begin{document}

\title{ \textbf{Bayesian Analysis of Glucose dynamics during the Oral Glucose Tolerance Test (OGTT)}}

\author{Hugo Flores-Arguedas, Marcos A. Capistr\'an\\ {\small Centro de Investigaci\'on en Matem\'aticas, A.C.} 
}

\maketitle

\begin{abstract}
This paper proposes a model that considers the action and timing of insulin and glucagon in glucose homeostasis after an oral stimulus. We use the Bayesian paradigm to infer kinetic rates, namely insulin and glucagon secretion, gastrointestinal emptying, and basal glucose concentration in blood. We identify two insulin scores related to glucose concentration in both blood and the gastrointestinal tract. The scores allow us to suggest a classification for individuals with impaired insulin sensitivity.
\end{abstract}

\section{Introduction}
This paper builds on the model of blood glucose dynamics during the Oral Glucose Tolerance Test (OGTT) of Kuschinski  \cite{kuschinski2016modeling} .  Our model considers the action and timing of insulin and glucagon in glucose homeostasis after an oral stimulus. We use the Bayesian paradigm to infer kinetic rates, namely insulin and glucagon secretion, gastrointestinal emptying, and basal glucose concentration in blood. We identify two insulin scores related to glucose concentration in both blood and the gastrointestinal tract, suggesting a classification for individuals with impaired insulin sensitivity.

Most tissues and organs in our body use glucose as an essential source of energy. The process of maintaining blood glucose at a steady state is called glucose homeostasis \cite{szablewski2011glucose}. Food ingestion, fasting, physical activity, or exercise constantly shift the blood glucose concentration away from equilibrium throughout the day. A low blood concentration of glucose can causes seizures, loss of consciousness, or even death. On the other hand, the long-lasting elevation of blood glucose concentrations can result in blindness, renal failure, vascular disease, and neuropathy. Therefore, blood glucose concentration in healthy individuals stays within narrow limits. Two hormones secreted by the pancreas: insulin and glucagon, are the primary regulators of blood glucose. The World Health Organization (WHO) refers to diabetes type 2 as a chronic and metabolic disease characterized by elevated blood glucose levels. Type 2 diabetes is more common in adults and occurs when the body becomes resistant to insulin or does not produce enough insulin. One standard test to screen for diabetes is the Oral Glucose Tolerance Test (OGTT). In this procedure, the patient must drink a sugary solution after eight-hour of fasting. Several blood samples are taken over the process, the first before consuming the sugary solution, the rest are taken afterward  to monitor the body's reaction to the glucose intake. A prediabetic state is characterized by an impaired fasting glucose level (IFG), an impaired blood glucose level at two hours after consuming the drink, referred to as impaired glucose tolerance (IGT), or both. Studies suggest that IFG may be associated with impaired insulin secretion while IGT is related to insulin resistance \cite{schianca2003significance}. OGTT based diabetes diagnostics suffers from uncertainties along the test \cite{salinari2011intestinal} on the amount of glucose absorbed and its absorption rate. A popular approach to describing glucose dynamics is compartmental models \cite{palumbo2013mathematical, makroglou2006mathematical}. The inference of model parameters from glucose observations is called an inverse problem.  The critical task of solving the inverse problem is to reliably predict quantities of interest in terms of the model´s variables and parameters. Of note, inverse problems are an imperfect path to knowledge due to errors in observation, modeling, and numerical simulation~\cite{oden2010computer}. The Bayesian paradigm allows solving the inverse problem producing predictions with quantified uncertainties. Under this approach, the solution of the inverse problem is a probability distribution for the parameters of interest \cite{kaipio2006statistical}, which describes multiple consistent scenarios to fit the data. In this work, we use the Bayesian approach to model and analyze OGTT data. Here the regressor comes from an ODE system that describes the glucose dynamics. We model insulin and glucagon secretion originated from blood glucose concentration. Also, we consider delays and hormonal effects from the gastrointestinal tract. \\ 

\noindent \textbf{Related Work} Research on glucose dynamics through compartmental models is quite common; see \cite{palumbo2013mathematical, makroglou2006mathematical} for reviews. Usually, the description level is related to the crucial features of the diagnostic tests involved. For the Oral Glucose Tolerance Test, one of the main features to be modeled is the gastrointestinal tract \cite{szablewski2011glucose, salinari2011intestinal, yokrattanasak2016simple}. Most models include the insulin effect on the glucose homeostasis process, include glucagon action is less common. Several works describe OGTT dynamics with a second-order differential equation \cite{ackerman1965model, wu2005case, zhang2016data, vargas2020estimation}. Authors in \cite{zhang2016data} have shown that a linear model may describe some postprandial glucose excursions. The process in the gastrointestinal tract has become relevant in the last years. Incretins are hormones secreted by the gut, which stimulate insulin secretion before blood glucose level increases. This effect is reduced or even vanished for type 2 diabetes patients, see \cite{nauck1986reduced,knop2007reduced}. Also, gut hormone secretion is low in the fasting state \cite{szablewski2011glucose}. Reactions of the body to the rise of the glucose level come with a delay, modeled by explicit delays in time \cite{saber2018mathematical,giang2008delay} or by extra compartments in the dynamical system \cite{kuschinski2016modeling, salinari2011intestinal, de2013routine}. In epidemiology, latency and infection periods are  usually described by an exponential distribution. Authors in \cite{lloyd2001realistic} argued the inadequacy of this distribution since this proposal overestimates shorter and longer durations of the phenomenon in the period. A gamma distribution may describe more realistic distributions. We can obtain this effect by subdividing the compartment into $n$ stages. Specifically, the distribution obtained is an Erlang distribution, a gamma distribution with a shape integer parameter \cite{champredon2018equivalence}. Inference problems look for determining the value of model parameters based on observed quantities of interest, like glucose, insulin, GLP-1, and/or GIP \cite{salinari2011intestinal, de2013routine}. 
For a complex model, often, several sources of data are required to infer the parameters. Using just glucose data limits the complexity and may lead to an identifiability problem of the parameters. Continuous Glucose Monitoring (CGM) provides information about the body reaction during a more extended period \cite{zhang2016data, goel2018minimal, eichenlaub2019minimal}. Authors in \cite{eichenlaub2019minimal} proposed a minimal model with CGM data and prove the structural identifiability using software packages.  Authors in \cite{wu2005case} proposed lifestyle adjustments for T2D subjects using glucose data for more than 4 hours each 30 min. The results show that the recovery time of post-prandial blood glucose level can be adjusted to 4 hours.  Bayesian approaches for estimate parameters of compartmental models were proposed in \cite{kuschinski2016modeling, vargas2020estimation}. Authors in \cite{vargas2020estimation} proposed a minimal model for glucose-insulin dynamics. They proposed interpreting the glucose-insulin system as a damped harmonic oscillator and suggested a classification based on two parameters: the peak of the curve and an average of rates.  Authors in \cite{kuschinski2019bayesian}  proposed a Bayesian experimental design for time's sample collection.  Since the inference results depend strongly on the quality of the data, an experimental design looks to improve the ability of the test to provide a more accurate result. Authors define a utility function for different time collections that allow determining an optimal among these candidates. For a Bayesian experimental design, this utility depends on posterior quantities. \\

\noindent \textbf{Contributions} The main contributions of this work are a model that includes: (i) A description of insulin secretion dependent on direct blood glucose level as well as gastrointestinal glucose. The incretin effect has a significant role in insulin secretion when the body reacts to   an oral stimulus. Separately, these two sources allow us to propose a classification between patients. These parameters may suggest possible alterations in healthy patients and determine misclassified subjects. We obtain those by a Bayesian inference that takes less than five minutes and five glucose measurements, making it a fast, reliable, and accessible tool. (ii) Delays of the body in the gastrointestinal tract and the endocrine system are modeled by gamma distributions. We introduce Erlang distributed periods in the hormonal dynamics for incretin, insulin, and glucagon. \\

\noindent \textbf{Limitations} We remark some limitations for this approach: (i) Since the glucagon dynamics turn on when the glucose level is lower than the basal level ($G_b$), for patients with all measurements higher than $G_b$, we do not have information about the parameter corresponding to this reaction. The consequence is that the posterior marginal for $\theta_2$ matches the prior marginal. Monitoring the glucose level for more than two hours may circumvent this situation.  (ii) The classification proposed is based on two parameters. Each is the product of two quantities with biological meaning. Nevertheless, we can not recover each action separately. Adding insulin data in the inference may allow recovery of each action parameter.  \\

\noindent The organization of the rest of this paper is as follows. In Section \ref{sec: modeling}, we present the ODE system that describes the biological situation for the glucose dynamics during OGTT.  This section includes the description of the Bayesian inference problem in Subsection \ref{sec: bayesian}. Our numerical results are present in Section \ref{sec: numerical}. Finally, we discuss our results and findings in Section \ref{sec: discussion}.

\section{Modeling of Glucose dynamics during OGTT}
\label{sec: modeling}
In this section, we describe the physiological situation in detail to justify our approach. The biological modeling on which we rely is given by the following ODE system 
\begin{eqnarray}
\label{eq: model_ogtt_glucose_base}
\dot{G} = & \lambda_1 L - \lambda_2 I + \lambda_3 V , \qquad &  G(0) = y_0  \\ 
\label{eq: model_ogtt_insulin_base}
\dot{I} = & \lambda_4 (G-G_b)^+ + \lambda_8 V - \lambda_5 I ,  \qquad &  I(0) = 0  \\
\label{eq: model_ogtt_glucagon_base}
\dot{L} = & \lambda_6 (G_b-G)^+-\lambda_7 L ,  \qquad &  L(0) = 0  \\
\dot{V} = & -\lambda_3 V ,  \qquad &  V(0) = V_0 
\label{eq: model_ogtt_digestive_base}
\end{eqnarray}
where $G$ denotes the glucose blood level ( in mg/dl), $I$ the insulin level above the basal insulin ( in $\mu$U/ml), $L$ the glucagon level above the basal glucagon ( in mg/dl), and $V$ the glucose from the sugary drink in the gastrointestinal tract ( in mg/dl). We propose a dynamical system centered in the basal glucose level denoted by $G_b$, which typically takes values in the range $[75-100]$ mg/dl, \cite{makroglou2006mathematical}. Above $G_b$, the dynamics are regulated mainly by insulin action, while below $G_b$, mainly due to glucagon. This system describes linear interactions for both cases. A linear description for the load path through the gastrointestinal system is usual, while nonlinear functions describe glucose appearance in blood \cite{de2013routine}. In this case, we decided to maintain the linearity assumption as in \cite{ackerman1965model}, which simplifies the biological phenomenon but reaches the objective of predicting the glucose level, a major quantity of interest for the inference \cite{oden2010computer}. Note that the complexity of models in \cite{salinari2011intestinal, de2013routine,  dalla2006system} comes with a diversity of observed data to validate it. In our case, we proposed the inference using just glucose data at five times, which is a limitation to increase the complexity. Also, the model's simplicity is not a limitation to obtain a good fit and reliable inference \cite{oden2010computer}.  Equation (\ref{eq: model_ogtt_glucose_base}) models the glucose dynamics, which decreases proportionally to the insulin level and increases according to two sources, first by the glucose of the sugary drink from the intestinal tract, second by the glucagon action. $\lambda_1$ and $\lambda_2$ are efficacy rates for the glucagon and the insulin, $\lambda_3$ is the rate of glucose absorption from the gastrointestinal tract. The dynamics of insulin and glucagon are very similar and depend on the basal glucose level of the body. The pancreas secretes insulin when the blood glucose level is higher than the basal glucose level on the body. Conversely, the pancreas secretes glucagon when the blood glucose level is lower than the basal glucose level on the body. We model this secretion action by a switch on the quantity $G-G_b$. Mathematically speaking, we use the positive part of $G-G_b$, denoted by $(G-G_b)^+$, and defined by
\begin{equation}
 (G-G_b)^+ =  \begin{cases} 
      G-G_b & G \geq G_b \\
      0 &  G \leq G_b 
   \end{cases} 
\end{equation} 
to model the secretion of insulin, conversely $(G_b-G)^+$ to the secretion of glucagon. $\lambda_4$ and $\lambda_6$ are secretion rates for the insulin and the glucagon due to blood glucose level, and $\lambda_5$ and $\lambda_7$ are its disintegration rates. $\lambda_8$ is the insulin secretion rate due to the glucose level in the gastrointestinal tract.  After a meal, insulin secretion may occur in two phases: an initial rapid release and a long-term release if glucose concentrations remain high \cite{szablewski2011glucose}. Hormones from the gut play a main role in insulin secretion on oral glucose consumption.  The second term on equation (\ref{eq: model_ogtt_insulin_base}) mimics the secretion due to a high concentration of glucose in the gastrointestinal tract as for example, the incretin effect. The Incretin effect is reduced or even vanished in diabetic patients \cite{knop2007reduced,nauck1986reduced}.  Therefore,  insulin secretion may be due to direct glucose blood level and/or gastrointestinal glucose level. The model defined by equations (\ref{eq: model_ogtt_glucose_base})- (\ref{eq: model_ogtt_digestive_base}) has eight parameters. Parameters $\lambda_5$ and $\lambda_7$ will be taken from the literature \cite{kuschinski2016modeling, duckworth1998insulin}. For the other six parameters, we will face an identifiability problem; that is, the structure of the model may lead to issues related to the uniquely estimation of parameters during the inference. To face this problem, we introduce scaled versions of insulin and glucagon. Note that by multiplying equation (\ref{eq: model_ogtt_insulin_base}) by a constant, we obtain a scaled insulin
\begin{equation}
\label{eq: scaled_insulin}
\lambda_I \dot{I} = \lambda_I \lambda_4 (G-G_b)^+ + \lambda_I \lambda_8 V - \lambda_I \lambda_5 I
\end{equation}
If we name $I_s = \lambda_I I$, we have $\dot{I_s} = \dot{\left(\lambda_I I\right)} = \lambda_I \dot{I}$. By substituying in equation (\ref{eq: scaled_insulin}), we obtain 
\begin{equation}
\label{eq: scaled_insulin_2}
\dot{I_s} = \theta (G-G_b)^+ + \gamma V- \lambda_5 I_s
\end{equation}
We will apply this idea to obtain a scaled insulin and glucagon by multiplying equation 
(\ref{eq: model_ogtt_insulin_base}) by $\lambda_2$ and equation (\ref{eq: model_ogtt_glucagon_base}) by $\lambda_1$. As we will explain later, we use only glucose observations for the inference, for which the loss of meaning for these scaled amounts do not represent a limitation. Finally we obtain the system 
\begin{eqnarray}
\label{eq: model2_ogtt_glucose_base}
\dot{G} = & L_1 - I_1 + \lambda_3 V \\ 
\label{eq: model2_ogtt_insulin_base}
\dot{I_1} = & \theta_1 (G-G_b)^+ + \theta_3 V - \lambda_5 I_1 \\
\label{eq: model2_ogtt_glucagon_base}
\dot{L_1} = & \theta_2 (G_b-G)^+-\lambda_7 L_1 \\
\dot{V} = & -\lambda_3 V 
\label{eq: model2_ogtt_digestive_base}
\end{eqnarray}
where $L_1 = \lambda_1 L, I_1 = \lambda_2 I, \theta_1 = \lambda_2 \lambda_4, \theta_3 = \lambda_2 \lambda_8$ and $\theta_2 = \lambda_1 \lambda_6$. Note that these substitutions allow us to decrease the number of parameters in the system, which will be helpful in the inference to fight the identifiability problem.  Table 1 summarizes information about the model's parameters given by equations (\ref{eq: model2_ogtt_glucose_base}) - (\ref{eq: model2_ogtt_digestive_base}).  For this model and the specific case when $G-G_b \geq 0$, we can show by the Similarity Transform Method \cite{vajda1989similarity} that $\theta_0$ is identifiable and $\theta_3$ is not, see section \ref{app: identifiability}.

Note that, for the cases of OGTT curves with all data greater than $G_b$, the term $(G_b-G)^+=0$ during most all the test, which causes a non-identifiability of $\theta_2$ and $G_b$. Let us recall that the unidentifiability  of parameters produces unrealistic values or high uncertainty around the estimated values of the parameters. A Bayesian approach deals with these problems by proposing a prior distribution on the parameters and approximating integrals from the MCMC, which produces less sensitivity on the results \cite{pillonetto2003numerical}.

\begin{table}[h!]
\begin{center}
\caption{Information on the model parameters }
\begin{tabular}{cccc} \hline
Parameter & Units & Description & Value \\\hline
 $\lambda_3$ & $hr^{-1}$ & GI Glucose Absorption  & Unknown \\
 $\theta_1$ & $hr^{-2}$ & Insulin Response to Blood Glucose Level & Unknown \\
 $G_b$ & mg/dl & Basal Glucose level & Unknown \\
 $\theta_3$ & $hr^{-2}$ &  Insulin response to GI Glucose Level & Unknown \\ 
 $\lambda_5$ & $hr^{-1}$ & Inverse of mean-life insulin clearance & 31 min, \cite{kuschinski2016modeling, duckworth1998insulin} \\ 
 $\theta_2$ & $hr^{-2}$ & Glucagon response to Blood Glucose  level & Unknown \\ 
 $\lambda_7$ & $hr^{-1}$ & Inverse of mean-life glucagon clearance & 31 min, \cite{kuschinski2016modeling, duckworth1998insulin} \\
 $V_0$ & mg/dl & Initial Glucose in Gastrointestinal Tract &  400,  Proposed  \\\hline 
\end{tabular}
\end{center}
\end{table}

Also, we would like to address the possible delays of the body reaction. Introducing explicit time delays into the ODE system is very common \cite{saber2018mathematical,giang2008delay}. The same effect may be obtained by introducing extra compartments. In epidemiology, these additional compartments model aspects as latency and infection periods with a more realistic distribution. This technique is known as an Erlang model \cite{lloyd2001realistic, champredon2018equivalence, capistran2021forecasting,  cassidy2020distributed}. Typically, people use the exponential distribution. Nevertheless, the exponential distribution overestimates the number of individuals whose duration of infection is shorter or longer than the mean \cite{lloyd2001realistic}. This approach subdivides stage $I$ (and/or $E$ ) into identical substages in a classical SEIR compartment model. This modeling matches the renewal approach \cite{champredon2018equivalence}, which considers the cohorts of infectious(exposed) individuals while the ODE approach considers each infectious(exposed) individual. In this work, we consider the same idea to model hormonal behavior during the OGTT test. That is, instead of considering an exponential decay of quantities $V_1, L_1$ and $I_1$, we introduce extra compartments to obtain a different and delayed behavior of these into the bloodstream. The main difference is illustrated in Figure \ref{fig: trace_plots} (a) for the digestive compartment. An exponential decay ($m=1$) gives higher weights to initial times. Note that the parameters $\lambda_5$ and $\lambda_7$ are the mean-life clearance of hormones insulin and glucagon. Under this approach, these parameters retain their meaning \cite{champredon2018equivalence}. We propose a modification of the model in \cite{capistran2016bayesian}, deduced before in equations (\ref{eq: model2_ogtt_glucose_base})-(\ref{eq: model2_ogtt_digestive_base}). This modification is given by 

\begin{eqnarray}
\label{eq: model_ogtt_glucose}
\dot{G} = & L_2 - I_2 + \theta_0 V_2 \\ 
\label{eq: model_ogtt_insulin1}
\dot{I}_1 = & \theta_1 (G-G_b)^+ + \theta_3 V_2 -2\lambda_5 I_1 \\
\dot{I}_2 = & 2\lambda_5 I_1-2\lambda_5 I_2 \\ 
\dot{L}_1 = & \theta_2 (G_b-G)^+-2\lambda_7 L_1 \\
\dot{L}_2 = & 2\lambda_7 L_{1}-2\lambda_7 L_2, \\ 
\dot{V}_1 = & -2 \theta_0 V_1 \\
\label{eq: model_ogtt_digestive2}
\dot{V}_2 = & 2\theta_0 V_{1}-2\theta_0 V_2 ,
\end{eqnarray}

\noindent where $\theta_0 = \lambda_3$. After ingestion, glucose is absorbed in the upper gastrointestinal tract, transported to the liver, and finally reaches the peripheral circulation \cite{dalla2006system}. Introducing a second compartment for $V$ allows us to model this behavior of glucose due to the sugary drink in the digestive system as in \cite{kuschinski2016modeling, flores2016bayesian}. Compartments $V_1$ and $V_2$ model the stomach and the small intestine. The second compartments for $I$ and $L$ allow us to model the delays due to the pancreas reaction \cite{saber2018mathematical}. These delays are a consequence of recurrent inhibitory dynamics \cite{eurich2002recurrent,mackey1984dynamics}. In recurrent inhibition, we can see how the activation of a quantity produces excitation in a second quantity that inhibits the first's activity. This phenomenon is present in the dynamics of glucose-insulin-glucagon. The specific case for glucose-insulin dynamics is known as a negative feedback loop \cite{de2008mathematical}. Following this approach, we can introduce extra compartments without new kinetic constants. This model is locally attractive to the unique equilibrium point $(G_b,0,0,0,0,0,0)$ for subjects with small $\theta_1$. We show details about the stability of the model in Subsection \ref{app: stability}. In the following subsection, we propose a regressor from the dynamical system in equations (\ref{eq: model_ogtt_glucose})-(\ref{eq: model_ogtt_digestive2}) to fit the glucose data provided by the OGTT. The inference process follows a Bayesian approach.

\subsection{Identifiability}
\label{app: identifiability}
In this section, we use the Similarity Transform Method \cite{vajda1989similarity} to show some results about the identifiability of parameters $\theta_0$ and $\theta_3$. Let us consider $X = G-G_b$ and the system for $X \geq 0$
\begin{equation}
\begin{pmatrix}
\dot{X} \\
\dot{I}_1 \\
\dot{L}_1 \\ 
\dot{V}_1 
\end{pmatrix}  =  \begin{pmatrix}
0 & -1 & 1 &  \theta_0 \\
\theta_1 & -\lambda_5  & 0 & \theta_3 \\
0 & 0 &  -\lambda_7  & 0 \\
0 & 0 & 0 & \theta_0  
\end{pmatrix} 
\begin{pmatrix}
X \\
I_1 \\
L_1 \\ 
V_1 
\end{pmatrix} 
=  A 
\begin{pmatrix}
X \\
I_1 \\
L_1 \\ 
V_1 
\end{pmatrix} 
\label{eq:SS1}
\end{equation}
subject to glucose observations 
\begin{equation}
\tilde{y} = X
\end{equation}
We need to find $T$, non singular, such that
\begin{equation}
A(\theta)T = T A(\tilde{\theta}) 
\end{equation}

\begin{equation}
\begin{pmatrix}
0 & -1 & 1 &  \theta_0 \\
\theta_1 & -\lambda_5  & 0 & \theta_3 \\
0 & 0 &  -\lambda_7  & 0 \\
0 & 0 & 0 & -\theta_0  
\end{pmatrix}
 \begin{pmatrix}
T_{11} & T_{12}  & T_{13} & T_{14} \\
T_{21} & T_{22}  & T_{23} & T_{24} \\
T_{31} & T_{32} & T_{33}  & T_{34} \\
T_{41} & T_{42} & T_{43} & T_{44}  
\end{pmatrix} =
 \begin{pmatrix}
T_{11} & T_{12}  & T_{13} & T_{14} \\
T_{21} & T_{22}  & T_{23} & T_{24} \\
T_{31} & T_{32} & T_{33}  & T_{34} \\
T_{41} & T_{42} & T_{43} & T_{44}  
\end{pmatrix}
 \begin{pmatrix}
0 & -1 & 1 &  \tilde{\theta_0} \\
\tilde{\theta_1} & -\lambda_5  & 0 & \tilde{\theta_3} \\
0 & 0 &  -\lambda_7  & 0 \\
0 & 0 & 0 & -\tilde{\theta_0}  
\end{pmatrix}
\end{equation}

\begin{equation}
\begin{pmatrix}
-T_{21} + T_{31} + \theta_0 T_{41} & -T_{22} + T_{32} + \theta_0 T_{42} & -T_{23} + T_{33} + \theta_0 T_{43} &  -T_{24} + T_{34} + \theta_0 T_{44} \\
\theta_1 T_{11} -\lambda_5 T_{21} + \theta_3 T_{41} & \theta_1 T_{12} -\lambda_5 T_{22} + \theta_3 T_{42}  & \theta_1 T_{13}-\lambda_5 T_{23} + \theta_3 T_{43} & \theta_1 T_{14} -\lambda_5 T_{24} + \theta_3 T_{44} \\
-\lambda_7 T_{31} & -\lambda_7 T_{32} & -\lambda_7 T_{33}  & -\lambda_7 T_{34} \\
-\theta_0 T_{41} & -\theta_0 T_{42} & -\theta_0 T_{43} & -\theta_0 T_{44}  
\end{pmatrix}
=
\end{equation}

\begin{equation}
 \begin{pmatrix}
T_{12} \tilde{\theta_1} & -T_{11}-\lambda_5 T_{12} & T_{11}-\lambda_7 T_{13} & \tilde{\theta_0}T_{11} +\tilde{\theta_3} T_{12}- \tilde{\theta_0}T_{14} \\
T_{22} \tilde{\theta_1} & -T_{21}-\lambda_5 T_{22}  & T_{21}-\lambda_7 T_{23} & \tilde{\theta_0}T_{21} +\tilde{\theta_3} T_{22}- \tilde{\theta_0}T_{24}\\
T_{32} \tilde{\theta_1} & -T_{31}-\lambda_5 T_{32} & T_{31}-\lambda_7 T_{33}  & \tilde{\theta_0}T_{31} +\tilde{\theta_3} T_{32}- \tilde{\theta_0}T_{34} \\
T_{42} \tilde{\theta_1} & -T_{41}-\lambda_5 T_{42} & T_{41}-\lambda_7 T_{43} & \tilde{\theta_0}T_{41} +\tilde{\theta_3} T_{42}- \tilde{\theta_0}T_{44} 
\end{pmatrix}
\nonumber
\end{equation}
After some computations, we can deduce that 
$$T_{31}=T_{32}=T_{34}=T_{41}=T_{42}=T_{43}=0$$
then 

\begin{equation}
\begin{pmatrix}
-T_{21}  & -T_{22}   & -T_{23} + T_{33}  &  -T_{24}  + \theta_0 T_{44} \\
\theta_1 T_{11} -\lambda_5 T_{21}  & \theta_1 T_{12} -\lambda_5 T_{22}   & \theta_1 T_{13}-\lambda_5 T_{23}  & \theta_1 T_{14} -\lambda_5 T_{24} + \theta_3 T_{44} \\
0 & 0 & -\lambda_7 T_{33}  & 0 \\
0 & 0 & 0 & -\theta_0 T_{44}  
\end{pmatrix}
=
\end{equation}

\begin{equation}
 \begin{pmatrix}
T_{12} \tilde{\theta_1} & -T_{11}-\lambda_5 T_{12} & T_{11}-\lambda_7 T_{13} & \tilde{\theta_0}T_{11} +\tilde{\theta_3} T_{12}- \tilde{\theta_0}T_{14} \\
T_{22} \tilde{\theta_1} & -T_{21}-\lambda_5 T_{22}  & T_{21}-\lambda_7 T_{23} & \tilde{\theta_0}T_{21} +\tilde{\theta_3} T_{22}- \tilde{\theta_0}T_{24}\\
0 & 0 & -\lambda_7 T_{33}  & 0 \\
0 & 0 & 0 & - \tilde{\theta_0}T_{44} 
\end{pmatrix}
\nonumber
\end{equation}
Since $T$ must be non singular, then $T_{44} \neq 0$ and $\tilde{\theta_0} = \theta_0$.  We have $\theta_0$ is identifiable.  Now, from the observation condition, we have $T_1 = (1,0,0,0)$, then 

\begin{equation}
\begin{pmatrix}
-T_{21}  & -T_{22}   & -T_{23} + T_{33}  &  -T_{24}  + \theta_0 T_{44} \\
\theta_1 -\lambda_5 T_{21}  &  -\lambda_5 T_{22}   & -\lambda_5 T_{23}  & -\lambda_5 T_{24} + \theta_3 T_{44} \\
0 & 0 & -\lambda_7 T_{33}  & 0 \\
0 & 0 & 0 & -\theta_0 T_{44}  
\end{pmatrix}
=
\end{equation}

\begin{equation}
 \begin{pmatrix}
0 & -1 & 1 & \tilde{\theta_0} \\
T_{22} \tilde{\theta_1} & -T_{21}-\lambda_5 T_{22}  & T_{21}-\lambda_7 T_{23} & \tilde{\theta_0}T_{21} +\tilde{\theta_3} T_{22}- \tilde{\theta_0}T_{24}\\
0 & 0 & -\lambda_7 T_{33}  & 0 \\
0 & 0 & 0 & - \tilde{\theta_0}T_{44} 
\end{pmatrix}
\nonumber
\end{equation}

Now, $T_{21} = 0$ and $T_{22} = 1$ and $\theta_1$ is identifiable,

\begin{equation}
\begin{pmatrix}
0  & -1   & -T_{23} + T_{33}  &  -T_{24}  + \theta_0 T_{44} \\
\theta_1  &  -\lambda_5   & -\lambda_5 T_{23}  & -\lambda_5 T_{24} + \theta_3 T_{44} \\
0 & 0 & -\lambda_7 T_{33}  & 0 \\
0 & 0 & 0 & -\theta_0 T_{44}  
\end{pmatrix}
=
 \begin{pmatrix}
0 & -1 & 1 & \tilde{\theta_0} \\
 \tilde{\theta_1} & -\lambda_5  & -\lambda_7 T_{23} & \tilde{\theta_3} - \tilde{\theta_0}T_{24}\\
0 & 0 & -\lambda_7 T_{33}  & 0 \\
0 & 0 & 0 & - \tilde{\theta_0}T_{44} 
\end{pmatrix}
\nonumber
\end{equation}

Then $T_{23} = 0$ and  $T_{33} = 1$

\begin{equation}
\begin{pmatrix}
0  & -1   & 1  &  -T_{24}  + \theta_0 T_{44} \\
\theta_1  &  -\lambda_5   & 0  & -\lambda_5 T_{24} + \theta_3 T_{44} \\
0 & 0 & -\lambda_7  & 0 \\
0 & 0 & 0 & -\theta_0 T_{44}  
\end{pmatrix}
=
 \begin{pmatrix}
0 & -1 & 1 & \tilde{\theta_0} \\
 \tilde{\theta_1} & -\lambda_5  & 0 & \tilde{\theta_3} - \tilde{\theta_0}T_{24}\\
0 & 0 & -\lambda_7   & 0 \\
0 & 0 & 0 & - \tilde{\theta_0}T_{44} 
\end{pmatrix}
\nonumber
\end{equation}
Finally, since $\tilde{\theta_0} = \theta_0$, we have that each
\begin{equation}
T =  \begin{pmatrix}
1 & 0  & 0 & 0 \\
0 & 1  & 0 & \theta_0(T_{44}-1) \\
0 & 0 & 1  & 0 \\
0 & 0 & 0 & T_{44}  
\end{pmatrix}
\end{equation}
with $T_{44} = \dfrac{\tilde{\theta_3} + \theta_0 (\theta_0 - \lambda_5)}{\theta_3 + \theta_0 (\theta_0 - \lambda_5)}$, is a similarity transform for the system and $\theta_3$ is unidentifiable.

\subsection{Stability}
\label{app: stability}

The process of maintaining blood glucose at a steady-state level is known as glucose homeostasis \cite{szablewski2011glucose}. In this case, we model this steady-state as the basal glucose level, denoted by $G_b$. The model proposed in equations (\ref{eq: model_ogtt_glucose})-(\ref{eq: model_ogtt_digestive2}) has an unique equilibrium point $(G_b,0,0,0,0,0,0)$. This model is piecewise linear, and we would like to demonstrate that every solution of the ODE system converges to this point which in practice means that under a perturbation or stimulus of the glucose level, the body looks for returning to balance. Note that we are not interested in the period of time it takes, even if for diabetic patients we expect greater times than for healthy patients.  Let us consider the change of variable $X = G-G_b$. The system given by equations (\ref{eq: model_ogtt_glucose})-(\ref{eq: model_ogtt_digestive2}) may be expressed as  
\begin{equation}
\begin{pmatrix}
\dot{X} \\
\dot{I}_1 \\
\dot{I}_2 \\
\dot{L}_1 \\ 
\dot{L}_2 \\
\dot{V}_1 \\  
\dot{V}_2
\end{pmatrix}  =  \begin{pmatrix}
0 & 0 & -1 & 0 & 1 & 0 & \theta_0 \\
\theta_1 & -2\lambda_5 & 0 & 0 & 0 & 0 & \theta_3 \\
0 & 2\lambda_5 & -2\lambda_5 & 0 & 0 & 0 & 0 \\
0 & 0 & 0 &  -2\lambda_7 & 0 & 0 & 0 \\
0 & 0 & 0 &  2\lambda_7 & -2\lambda_7 & 0 & 0\\
0 & 0 & 0 & 0 & 0 & -2\theta_0 & 0 \\
0 & 0 & 0 & 0 & 0 & 2\theta_0 & -2\theta_0 
\end{pmatrix} 
\begin{pmatrix}
X \\
I_1 \\
I_2 \\
L_1 \\ 
L_2 \\
V_1 \\  
V_2
\end{pmatrix} 
\label{eq:SS1}
\end{equation}
for $x\geq 0$, and by 
\begin{equation}
\begin{pmatrix}
\dot{X} \\
\dot{I}_1 \\
\dot{I}_2 \\
\dot{L}_1 \\ 
\dot{L}_2 \\
\dot{V}_1 \\  
\dot{V}_2
\end{pmatrix}  =  \begin{pmatrix}
0 & 0 & -1 & 0 & 1 & 0 & \theta_0 \\
0 & -2\lambda_5 & 0 & 0 & 0 & 0 & \theta_3 \\
0 & 2\lambda_5 & -2\lambda_5 & 0 & 0 & 0 & 0 \\
-\theta_2 & 0 & 0 &  -2\lambda_7 & 0 & 0 & 0 \\
0 & 0 & 0 &  2\lambda_7 & -2\lambda_7 & 0 & 0\\
0 & 0 & 0 & 0 & 0 & -2\theta_0 & 0 \\
0 & 0 & 0 & 0 & 0 & 2\theta_0 & -2\theta_0 
\end{pmatrix} 
\begin{pmatrix}
X \\
I_1 \\
I_2 \\
L_1 \\ 
L_2 \\
V_1 \\  
V_2
\end{pmatrix} 
\label{eq:SS2}
\end{equation}
for $x<0$. The unique equilibrium point of this piece-wise linear systems is $(0,0,0,0,0,0,0)$. According to \cite{liberzon2003switching}, the switched linear system given by equations 
(\ref{eq:SS1}) and (\ref{eq:SS2}) is Globally Uniform Exponential Stable (GUES) iff it is locally attractive for every switching signal (see Theorem 2.4). Let us recall that the local attractivity property means that all trajectories starting in some neighborhood of the origin converge to the origin. 
For $x\geq 0$, we can convert system in Equation (\ref{eq:SS1}) in the third order linear equation 
\begin{equation}
\label{eq: third_harmonic_osc}
x^{(3)} + 4 \lambda_5 \ddot{x} + 4 \lambda^2_5 \dot{x} + 2\lambda_5 \theta_1 x = f(t) 
\end{equation}
with $f(t) \to 0$. Associated to equation (\ref{eq: third_harmonic_osc}), we have a third order polynomial
\begin{equation}
\label{eq: third_poly}
at^3 + b t^2 + ct + d = 0 
\end{equation}
 which roots determine the behavior of its solutions. Let us recall that the discriminant of a depressed cubic $t^3 + pt + q = 0$ is $\Delta = -(4p^3 + 27q^2)$. For equation (\ref{eq: third_poly}), after a change of variable to obtain it depressed form, we have
\begin{equation}
\Delta = -(4p^3 + 27q^2) = -4\lambda^2_5 \theta_1 (-16 \lambda^2_5 + 27 \theta_1)
\end{equation}
Recall that for $\Delta<0$, the equation has one real root and two non-real complex conjugate, while for $\Delta>0$, three real roots. Note that the value $\theta_1 = \dfrac{16}{27}\lambda^2_5$ satisfies $\Delta=0$.
 
\noindent For $\theta_1<\dfrac{16}{27}\lambda^2_5$, $\Delta>0$ is satisfied. In this case is very simple to verify that the three real roots are negative. Consider the polynomial $G(t) = t^3 + 4 \lambda_5 t^2 + 4 \lambda^2_5 t + 2\lambda_5 \theta_1 $, which satisfies:
\begin{equation}
G(-2/3 \lambda_5)= \quad 2\lambda_5(\theta_1 -16/27 \lambda_5^2) \quad < \quad 0 \quad <  2\lambda_5 \theta_1\quad =G(0)
\end{equation}
that means, there exist a root $x_1 \in ]-2/3 \lambda_5,0[$ of $G(t)$. Analogously, we can find $x_2 \in ]-2 \lambda_5, -2/3 \lambda_5[$ and $x_3 \in ]-4 \lambda_5, -2 \lambda_5[$ roots of $G(t)$. Since all are negative, the local attractivity property is satisfied in this case.

\noindent For $\theta_1>\dfrac{16}{27}\lambda^2_5$, $\Delta<0$ is satisfied. In this case, the roots of $G(t)$ are given by 
\begin{equation}
x_k = -\dfrac{1}{3a}\left( b + \xi^k C + \dfrac{\Delta_0}{\xi^kC}\right), \qquad k = 0,1,2
\end{equation}
with $\xi = \dfrac{-1 + \sqrt{3}i}{2}$, $C = \left(\dfrac{\Delta_1 + \sqrt{\Delta_1^2 -4 \Delta_0^3}}{2} \right)^{1/3}$, $\Delta_0 = b^2-3ac$ and $\Delta_1 = 2b^3 - 9abc + 27a^2d$. Since $a = 1,  b = 4\lambda_5, c = 4\lambda^2_5, d =2\lambda_5 \theta_1$, we have
\begin{equation}
\Delta_0 = 16\lambda_5^2-12\lambda_5^2 = 4\lambda_5^2 , \qquad \Delta_1 = -16 \lambda_5^3 + 54 \lambda_5 \theta_1 = 2 \lambda_5 (-8 \lambda_5^2 + 27 \theta_1 ) 
\end{equation}
then 
$$\Delta_1^2 -4 \Delta_0^3 = \left(-2\lambda_5(8\lambda^2_5 - 27\theta_1)\right)^2-4\cdot 64 \lambda_5^6=4 \lambda_5^2 \left( (8\lambda^2_5 - 27\theta_1)^2 - 64 \lambda_5^4\right)$$
$$ = 4 \lambda_5^2 \left( -2\cdot 8\lambda^2_5\cdot 27\theta_1 + 27^2\theta_1^2 \right) = 4 \lambda_5^2 \cdot 27\theta_1\left( -16\lambda^2_5  + 27\theta_1 \right)$$
Since $\theta_1>\dfrac{16}{27}\lambda^2_5$ then $C>0$, and therefore $x_0 = -\dfrac{1}{3a}\left( b + C + \dfrac{\Delta_0}{C}\right)$ is real and negative.  Now, $x_1 = -\dfrac{1}{3a}\left( b + \xi C + \dfrac{\Delta_0}{\xi C}\right)= -\dfrac{1}{3a}\left( b + \xi C + \xi^2 \dfrac{\Delta_0}{ C}\right) =$ \\\
$= -\dfrac{1}{3a}\left( b -\dfrac{1}{2} C - \dfrac{\Delta_0}{2 C} + \dfrac{\sqrt{3}}{2}\ i \left(C - \dfrac{\Delta_0}{ C} \right) \right)$. Its real part is $-\dfrac{1}{3a}\left( b -\dfrac{1}{2} C - \dfrac{\Delta_0}{2 C}  \right)$.  We are interested in determining under what conditions $\left( b -\dfrac{1}{2} C - \dfrac{\Delta_0}{2 C}  \right)>0$. A straightforward computation leads to the condition 
\begin{equation}
(4-2\sqrt{3})\lambda_5 < C < (4+2\sqrt{3})\lambda_5.
\end{equation}
This condition establish an upper bound for the value of $\theta_1$. In practice, this condition is satisfy for values of $\theta_1 <29$. A similar analysis is valid for $x<0$ and $\theta_2 <29$. 

\subsection{Bayesian Formulation for the Inference Problem}
\label{sec: bayesian}
In this section, we consider the inverse problem of determining the posterior distribution for the parameter $\Theta = (\theta_0, \theta_1, \theta_2, G_b, \theta_3)$. Under the Bayesian approach, the solution of the inverse problem is a probability distribution conditioned on the information of the data. We have glucose measurements at five times $t_i=0.0, 0.5,1.0, 1.5$ and $2.0$ hours. We assume that the data $\textbf{y}$ follow the noise model
\begin{equation}
\textbf{y}_i = \mathcal{G}(\Theta)_i + \eta_i
\label{eq:fit_regressor}
\end{equation}
where $\mathcal{G}(\Theta)_i = G(t_i,\Theta)$  and $G$ is the solution for the glucose on the ODE system given in equation (\ref{eq: model_ogtt_glucose}) and $\eta_i \sim \mathcal{N}(0, \sigma^2)$.  The parameters $\lambda_5, \lambda_7$ are determined from previous works \cite{kuschinski2016modeling, flores2016bayesian}. We propose $\sigma=5 $ mg/dl as in \cite{kuschinski2016modeling,vargas2020estimation}.

\noindent Our prior knowledge about the parameters are that $\theta_i>0$ and the basal level of glucose is in the range of $[75,100]$ mg/dl  \cite{makroglou2006mathematical}. We assumed independence on the prior parameters, that is  
\begin{equation}
\pi_0(\Theta) = \pi^0_0(\theta_0)\pi^1_0(\theta_1)\pi^2_0(\theta_2)\pi^3_0(G_b)\pi^4_0(\theta_3) 
\end{equation} 
We propose gamma distribution for each parameter. Le us recall that if $Z \sim \Gamma (\alpha, \beta)$ then $\mathbb{E}[Z]=\alpha / \beta$ and $\mathbb{V}ar[Z]=\alpha /{\beta^2}$. Our priors proposal are given by:
\begin{eqnarray}
\theta_{0} & \sim \Gamma (2 ,1 ) \nonumber \\
\theta_{1}  & \sim \Gamma (10 ,1 ) \nonumber \\
\theta_{2} & \sim \Gamma (10 ,1 ) \nonumber \\
\theta_{3}  & \sim \Gamma (10 ,1 ) \nonumber \\
G_b & \sim \Gamma (90^2/20 ,90/20 ) .  
\label{eq:prior_distributions_ogtt}
\end{eqnarray}
The prior distribution for the parameter $\theta_0$ corresponding to the gastrointestinal dynamic is truncated. From simulations, we propose values greater than 0.5 to avoid almost constant trajectories for the glucose. For $G_b$, we consider a prior with mean $90$ mg/dl and variance $20$ mg/dl. The priors for $\theta_1, \theta_2$ and $\theta_3$ are less informative with same mean and variance in 10. Having the likelihood $\pi(\textbf{y} | \Theta)$ and the prior $\pi_0(\Theta)$, and since we are working in finite dimensions, Bayes theorem ensures the posterior distribution $\pi(\Theta|\textbf{y})$ existence and
\begin{equation}
\label{eq: post_distr}
\pi(\Theta|\textbf{y}) \propto \pi(\textbf{y} | \Theta) \pi_0(\Theta).
\end{equation}
Since we are using a numerical regressor in equation (\ref{eq:fit_regressor}), the posterior distribution has no closed-form, and we need to generate samples of it. Note from equation (\ref{eq: post_distr}) that the posterior distribution is known up to a constant of proportionality. Markov Chain Monte Carlo (MCMC) allows obtaining samples from a distribution under this condition. Let us recall that the chain is a sequence of random variables obtained from a transition kernel and distributed according to a stationary probability distribution. To illustrate the results, we consider estimators for this distribution. The more popular estimators are the conditional mean (CM), which is the mean of the posterior distribution; the conditional median, which is the median of the posterior distribution; or the maximum a posteriori (MAP), which is the value of the parameters where the posterior reaches its maximum value. These estimators are computed based on the samples generated from the MCMC. To generate samples, we used the t-walk \cite{christen2010general}. We perform 10000 iterations for each patient with a burnin of 1000. To assess the efficiency of the sampler, we compute the integrated autocorrelation time (IAT). IAT is the number of steps required for the chain to produce independent samples. Authors in \cite{roberts2001optimal} argue that $IAT/n$, where $n$ is the number of parameters, is a measure of efficiency among MCMC samplers. Let us recall that the Monte Carlo error introduced in the approximations of integrals is proportional to $\sqrt{IAT/N}$, where $N$ is the number of samples from the posterior distribution \cite{hogg2018data}. That means we are looking for small values of the IAT to have efficiency. We show our results in the following subsection.

\section{Numerical Results}
\label{sec: numerical}

\begin{figure}[h!]
\begin{subfigure}[h]{0.48\textwidth}
 \includegraphics[width=\textwidth]{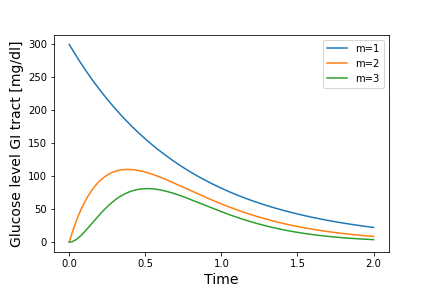}\caption{ }
\end{subfigure} 
\begin{subfigure}[h]{0.48\textwidth}
 \includegraphics[width=\textwidth]{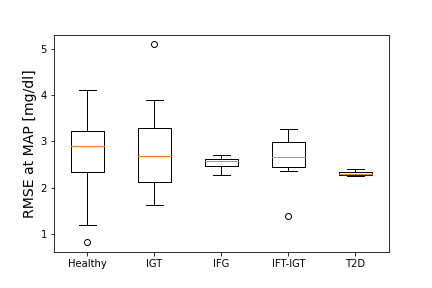}\caption{ }
\end{subfigure} 
\begin{subfigure}[h]{0.48\textwidth}
 \includegraphics[width=\textwidth]{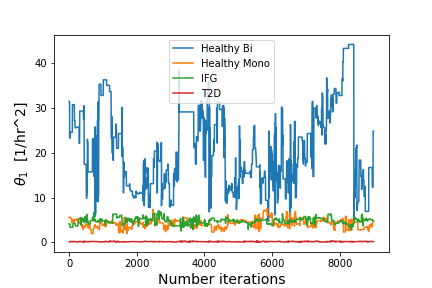}\caption{ }
\end{subfigure} 
 \begin{subfigure}[h]{0.48\textwidth} 
 \includegraphics[width=\textwidth]{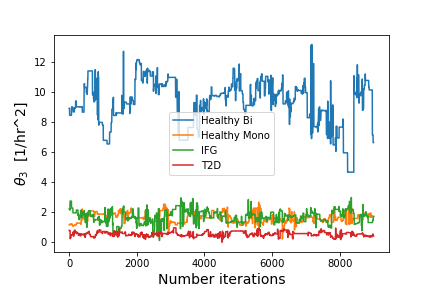}\caption{ }
 \end{subfigure}
\caption{(a) Erlang's distribution effects for $1,2$ and $3$ compartments in the gastrointestinal (GI) tract. (b) Boxplots of the RMSE at the MAP for each patient's category. Note that the results are consistent with the standard deviation proposed on the likelihood ($\sigma = 5$). Trace plots for parameters $\theta_1$ in (c) and $\theta_3$ in (d) for four patients with different characteristics: healthy biphasic, healthy monophasic, IGT and T2D.  Their corresponding  IAT/5 are 38, 56, 31,10.}
\label{fig: trace_plots}
\end{figure}

\begin{figure}[h!]
\begin{subfigure}[h]{0.32\textwidth}
 \includegraphics[width=\textwidth]{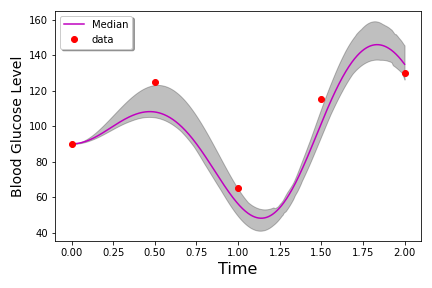}\caption{ }
\end{subfigure} 
\begin{subfigure}[h]{0.32\textwidth}
 \includegraphics[width=\textwidth]{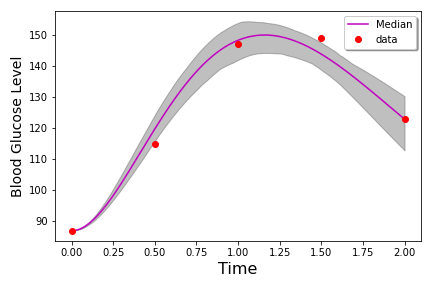}\caption{ }
\end{subfigure} 
\begin{subfigure}[h]{0.32\textwidth}
 \includegraphics[width=\textwidth]{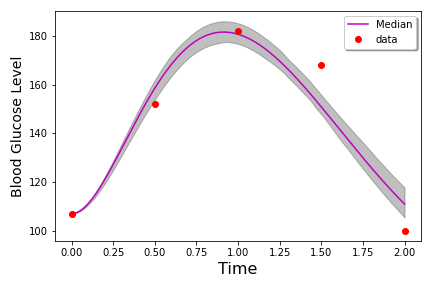}\caption{ }
\end{subfigure} 
\begin{subfigure}[h]{0.32\textwidth}
 \includegraphics[width=\textwidth]{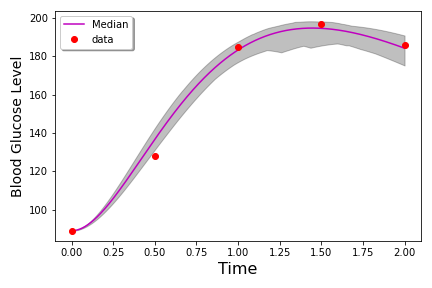}\caption{ }
\end{subfigure} 
 \begin{subfigure}[h]{0.32\textwidth} 
 \includegraphics[width=\textwidth]{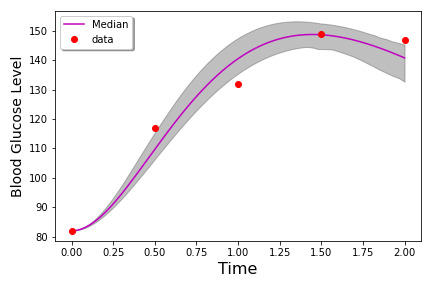}\caption{ }
 \end{subfigure}
  \begin{subfigure}[h]{0.32\textwidth} 
 \includegraphics[width=\textwidth]{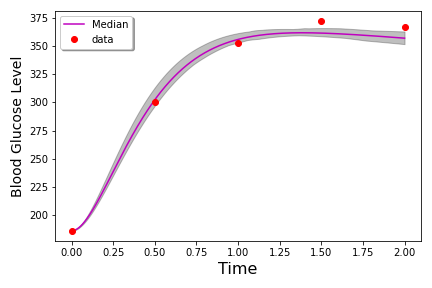}\caption{ }
 \end{subfigure}
\caption{Glucose data (red points), the median of the posterior distribution for each patient and 200 trajectories of the posterior distribution to illustrate the uncertainty (grey area): (a) Healthy Patient with Biphasic curve, (b) Healthy Patient with Monophasic curve, (c) Patient with IFG , (d) Patient with IGT (high glucose level at $t=2$), (e) Patient with IGT (low glucose level at $t=2$), (f) Diabetic patient.}
\label{fig: FIT_glucoseUQ}
\end{figure}

In this subsection, we show the results of the inference. We perform a MCMC using t-walk. Data were collected from 2012 to 2019 in the General Hospital of Mexico \textit{Dr. Eduardo Liceaga} under the research \textit{Identificación de factores que predisponen a la diabetes mellitus tipo 2 en sujetos normoglicémicos con historia familiar de diabetes y su relación con obesidad}. These data were previously used in \cite{vargas2020estimation}. All participants signed a consent. A first measurement is taken at fasting, after eight hours of fasting. After that, the patient has five minutes to consume a drink with 75 gr. of dextrose.  Four different measurements of glucose are taken at 30, 60, 90, and 120 minutes. We have data from 80 female patients classified as follows:
\begin{enumerate}
\item 51 healthy patients (H).
\item 4 patients with Impaired Fasting Glucose (IFG): Fasting blood glucose level $\ge 100$.
\item 15 patients with Impaired Glucose Tolerance (IGT): Blood glucose level $\ge 140$ at $t=2$.
\item 7 patients with IFG and IGT (IFG-IGT: both alterations).
\item 3 patients with Diabetes Mellitus 2 (T2D): Fasting blood glucose level $\ge 126$ and blood glucose level $\ge 200$ at $t=2$.
\end{enumerate}
Since only a fraction of the ingested glucose appears in the blood \cite{dalla2005insulin}, we propose a value of $V_0 = 400$ based on the maximum value of the data set (375 mg/dl). The initial condition for the glucose level $G(0)$ is equal to the fasting observation $y_0$.
Figure \ref{fig: trace_plots} (a) illustrates Erlang's distribution effects for $1,2$, and $3$ compartments in the digestive tract.  The case $m=1$ corresponds to an exponential distribution. This case may describe the process of the glucose level in the stomach. For $m=2$ and $3$, the quantity of glucose at $t=0$ is 0 and may describe different parts of the intestine. We proposed the case $m=2$ based on simulation results. Figure \ref{fig: trace_plots} (b) shows the RMSE at the MAP estimate for each category of patients in boxplots. The results are consistent with the standard deviation proposed on the likelihood ($\sigma = 5$). In Figure \ref{fig: trace_plots} (c)-(d), we show trace plots for the parameters $\theta_1$ and $\theta_3$ for several condition patients. These results illustrate the chain values without the burnin. In Figure \ref{fig: FIT_glucoseUQ}, we show the fit to the data for several patients. We include (in grey) trajectories generated from posterior samples and the curve generated by the posterior median. The CM and the MAP estimates generate very similar trajectories.

\begin{figure}[h!]
\begin{subfigure}[h]{0.6\textwidth}
 \includegraphics[width=\textwidth]{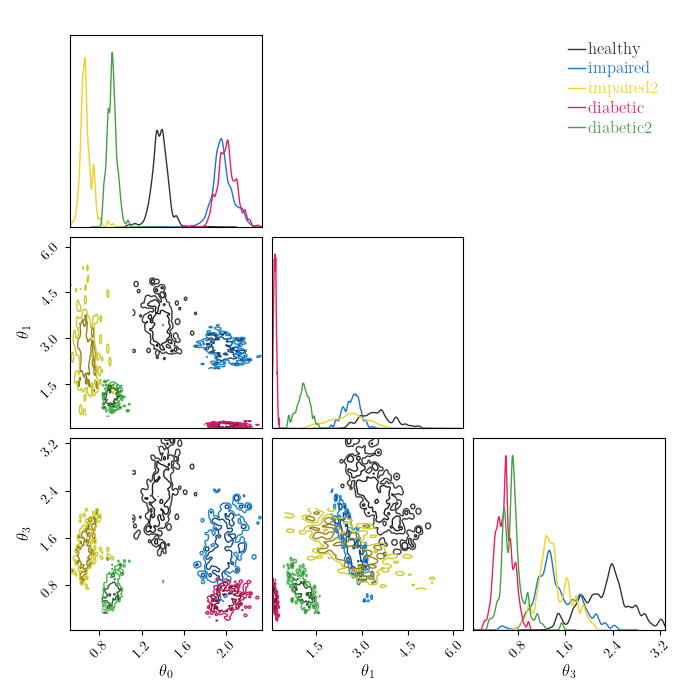}\caption{ }
\end{subfigure} 
\begin{subfigure}[h]{0.35\textwidth}
 \includegraphics[width=\textwidth]{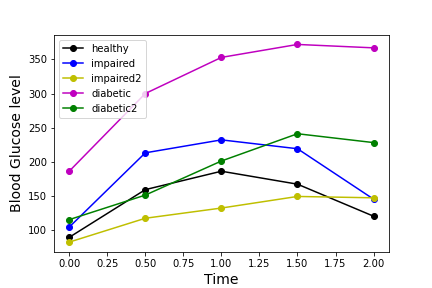}\caption{ }
\end{subfigure} 
\caption{ (a) Chain values of the MCMC for parameters $\theta_0, \theta_1$, and $\theta_3$ for two diabetic patients, two patients with impaired Glucose Tolerance and a Healthy patient. Gastric emptying parameter $\theta_0$ allow us to recognize initial slope during the test. (b) Glucose level measurements from patients in (a).}
\label{fig: Classification_chains_data}
\end{figure}

From the inference information, we would like to explore the values of the parameters that describe different situations on patients. In Figure \ref{fig: Classification_chains_data} (a), we show chain values for five different scenarios. In Figure \ref{fig: Classification_chains_data} (b), we show the corresponding glucose measurements. Note that parameters $\theta_1$ and $\theta_3$ allow us to recognize low secretion insulin levels. The gastrointestinal emptying parameter $\theta_0$ allows us to identify the initial slope during the test. That is, $\theta_0$ determines the increase from level at fasting and level at $t= 30$ minutes. Two patients with low gastrointestinal emptying labeled as impaired2 and diabetic2  present higher values close to two hours instead of at the beginning.  Note that this parameter does not allow us to recognize anomalies since, for two diabetic patients, the parameter may take very different values. 

In Figure \ref{fig: Posterior_var} (a)-(c), we show the posterior variance for the parameters $\theta_0, \theta_1$ and $\theta_3$ inferred from the MCMC. Note that for $\theta_1$, there are some healthy outliers with high variance, consequence of the unindentifiability problem, see details in Appendix \ref{app: inf_results}.
In Figure \ref{fig: Classification} (a), we show the values of the MAP estimate for parameters $\theta_1$ and $\theta_3$ inferred. We plotted $1/\theta_3$ against $1/\theta_1$. A high value for $1/\theta_1$ may be produced by a high glucose level at time 30 minutes whilst a high value for $1/\theta_3$ may be produced by a high glucose level at time 1 hour and 30 minutes. Patients diagnosed with IFG are considered healthy in the classification based on insulin secretion indicators. Its condition is not a consequence a the oral stimulus. We performed a linear Support Vector Machine (SVM) algorithm using the package scikit-learn in python \cite{pedregosa2011scikit}. An SVM is a supervised machine learning algorithm for classification, which has become popular in biological applications \cite{noble2006support}. In this case, we are looking for a separating line between two different classes, healthy patients and patients with an impaired condition. This hyperplane is optimal in the sense that it results from an optimization problem. Since naturally there is a transition stage from a healthy patient and a diabetic patient, we performed this algorithm for all the quantiles from 10 to 90 to include the uncertainty of our simulations in the classification.  In Figure \ref{fig: Classification} (b), we show the glucose data, the fit, and the uncertainty corresponding to the patient with IGT inside the healthy zone in  (a) (cyan star at left down corner). For this case, the parameter $G_b$ is identifiable, see Figure \ref{fig: Posterior_var} (d). We show results of the inference for this patient in Table 2. Our results suggested that this subject may be misclassified as IGT instead of healthy.

\begin{figure}[h!]
\begin{subfigure}[h]{0.48\textwidth}
 \includegraphics[width=\textwidth]{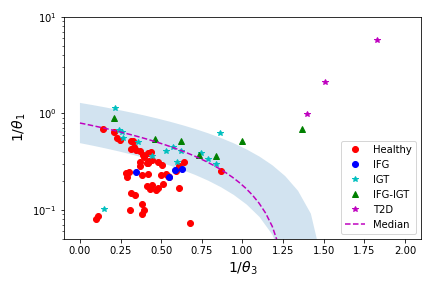}\caption{ }
\end{subfigure} 
\begin{subfigure}[h]{0.48\textwidth}
 \includegraphics[width=\textwidth]{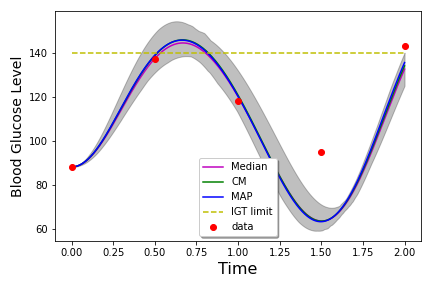}\caption{ }
\end{subfigure} 
\caption{(a) Plot $1/\theta_3$ against $1/\theta_1$ for the MAP estimate values: a linear SVM allow us to produce a classification between healthy and unhealthy patients. This algorithm performed from 10 to 90 quantile illustrate a transition stage between healthy and diabetic patients including the uncertainty of our simulations,  (b) Glucose data and fitted trajectories from an IGT patient with healthy indicators values (cyan star at left down corner in (a)). Note that the impaired glucose at $t=2$ is due to the glucagon secretion effect. This last measurement is almost in the limit of the IGT and the healthy classes.}
\label{fig: Classification}
\end{figure}

In Figure \ref{fig: Posterior_var} (a)-(c), we show some results about the inference as the posterior variance of parameters $\theta_0, \theta_1$ and $\theta_3$. Since parameters $\theta_2$ and $G_b$ are not practical identifiable for most of the subjects, we omit this information. We show inference results for a possible misclassified IGT patient in Figure \ref{fig: Posterior_var} (d) and Table 2. 
\begin{figure}[h!]
\begin{subfigure}[h]{0.48\textwidth}
 \includegraphics[width=\textwidth]{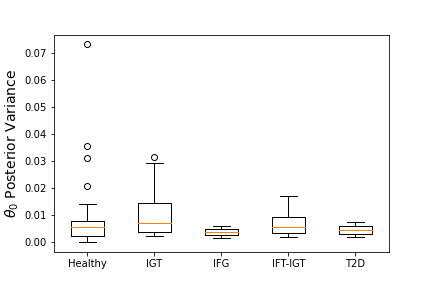}\caption{ }
\end{subfigure} 
\begin{subfigure}[h]{0.48\textwidth}
 \includegraphics[width=\textwidth]{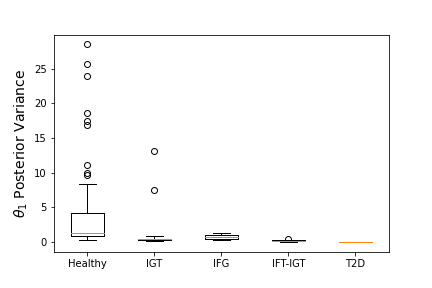}\caption{ }
\end{subfigure} 
\begin{subfigure}[h]{0.48\textwidth}
 \includegraphics[width=\textwidth]{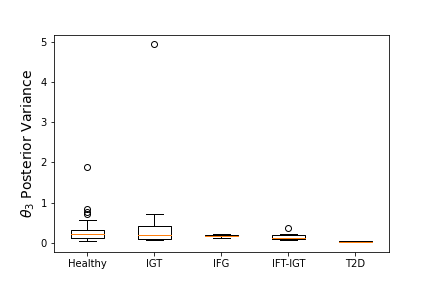}\caption{ }
\end{subfigure} 
\begin{subfigure}[h]{0.48\textwidth}
 \includegraphics[width=\textwidth]{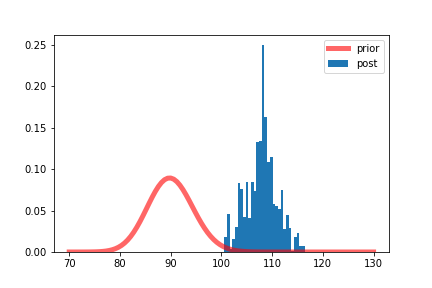}\caption{ }
\end{subfigure} 
\caption{(a)-(c) Posterior Variance for $\theta_0, \theta_1$ and $\theta_3$ by category. Note that for healthy patients, we can find outliers with high variance, probably related with the unidentifiability of these parameters. (d) Prior and posterior marginals of the parameter $G_b$ for misclassified IGT patient in Figure \ref{fig: Classification} (b)}
\label{fig: Posterior_var}
\end{figure}

\begin{table}[h!]
\begin{center}
\caption{Statistics of misclassified IGT patient in Figure \ref{fig: Classification} (b)}
\begin{tabular}{cccccc} \hline
Indicator/Parameter & $\theta_0$ & $\theta_1$ & $\theta_2$ & $G_b$ & $\theta_3$ \\\hline
MAP & 0.96 &  9.77 & 45.07 & 108.9 & 6.77 \\
CM &  0.96 &  11.55 & 45.11 & 107.9 & 6.2 \\
Median &  0.93 &  10.77 & 44.6 & 108.1 & 6.44 \\
 std &  0.13 &  3.6 & 4.5 & 3.06 & 0.84 \\\hline 
\end{tabular}
\end{center}
\end{table}

\section{Discussion and conclusions}
\label{sec: discussion}

In this work, we propose a Bayesian approach to analyze OGTT data. The modeling includes two insulin indicators, one related to blood glucose level, the other with the glucose level in the gastrointestinal tract.  These indicators describe insulin dynamics due to oral stimulus. Figure \ref{fig: Classification} (c) illustrates a possible classification for patients. Now, we discuss its scope and limitations. Patients diagnosed with IFG have an impaired value at fasting state and unimpaired values at two hours. Their insulin indicators classify these patients as healthy because their situation is not a consequence of the oral stimulus. A patient with IGT may have \textit{healthy} values for its insulin indicators, as shown in Figure \ref{fig: Classification} (a) (cyan star in the left down corner). From its glucose values (Figure \ref{fig: Classification} (b)), we can deduce that its situation is a consequence of impaired glucagon secretion. Also, note that the glucose measurement at $t = 2$ is almost on the limit of a healthy condition. The insulin indicators allow recognizing this possible misclassification. Note that the detection of these cases is conditioned to have more data than fasting and 2h level sample. Finally, the classification places some healthy patients in the transition zone. This situation may result from a high glucose level at 30 minutes, for $\theta_1$, or at 1 hour and 30 minutes for $\theta_3$, or a measurement close to 140 at 2 hours. This situation may be the beginning of an anomaly. Control and follow-up of these patients for several months or years may confirm it. The scores presented here have physiological meaning, discriminate between healthy and diabetic patients, and are accessible since the inference takes less than five minutes and the data needed to perform it are just from glucose measurements. From our results, it seems to us that the classification in healthy, IFG, IGT, IFG-IGT, and T2D could be modified to recognize more specific details about the cause of an impaired condition. IGT subjects may suffer from diminished glucagon suppression and an impaired glucose level at 2 h may be caused by this irregularity \cite{henkel2005impact}. Recognize the real causes of an impaired condition leads to different therapies and savings of health care resources. Also, early detection of anomalies and/or high risk of developing T2D allows for implementing healthy habits therapeutic strategies and postponing medication. The model described here is very simple but incorporates all essential elements of the physiological situation. As we mentioned before, the lack of complexity corresponds to the limited diversity of data. In this case, just with five glucose measurements, a very basic observation scenario, we were allowed to propose a classification. This classification enables recognizing possible misclassified patients and determining a transition zone, which may represent the risk of developing anomalies for healthy patients. Future work contemplates how to adapt the model to incorporate insulin and glucagon data. This information may provide possible explanations for an impaired condition.

\noindent{\large\bf Acknowledgments}{ \\ We would like to thank J. Andr\'es Christen for his help and comments and Antonio Capella for valuable comments.  Also, we would like to thank Adriana Monroy for the data and comments. HF-A was partially found by RDCOMM grant from J. Andr\'es Christen. The authors are partially founded by CONACYT CB-2016-01-284451 grant.}

\bibliography{references.bib}{}
\bibliographystyle{unsrt}
\nocite{*}

\end{document}